\documentclass[12pt,english,aps,manuscript]{revtex4}
\usepackage[T1]{fontenc}
\usepackage[latin9]{inputenc}
\usepackage{color}
\usepackage{float}
\usepackage{amsmath}
\usepackage{amssymb}
\usepackage{graphicx}

\makeatletter

\newcommand*\LyXThinSpace{\,\hspace{0pt}}
\newcommand{\noun}[1]{\textsc{#1}}

\@ifundefined{textcolor}{}
{%
 \definecolor{BLACK}{gray}{0}
 \definecolor{WHITE}{gray}{1}
 \definecolor{RED}{rgb}{1,0,0}
 \definecolor{GREEN}{rgb}{0,1,0}
 \definecolor{BLUE}{rgb}{0,0,1}
 \definecolor{CYAN}{cmyk}{1,0,0,0}
 \definecolor{MAGENTA}{cmyk}{0,1,0,0}
 \definecolor{YELLOW}{cmyk}{0,0,1,0}
}

\makeatother

\usepackage{babel}
\begin{document}

\title{Modulation of spin-orbit coupled Bose-Einstein condensates: analytical
characterization of acceleration-induced transitions between energy
bands}

\author{J.M. Gomez Llorente and J. Plata}

\address{Departamento de F\'{\i}sica, Universidad de La Laguna,\\
 La Laguna E38204, Tenerife, Spain.}
\begin{abstract}
The effects of modulating spin-orbit coupled Bose-Einstein condensates
are analytically studied. A sinusoidal driving of the coupling amplitude
is shown to induce significant changes in the energy bands and in
the associated spin-momentum locking. Moreover, in agreement with
recent experimental results, gravitational acceleration of the modulated
system is found to generate transitions between the modified energy
bands. The applicability of the Landau-Zener (LZ) model to the understanding
of the experimental findings is rigorously traced. Through a sequence
of unitary transformations and the reduction to the spin space, the
modulated Hamiltonian, with the gravitational potential incorporated,
is shown to correspond to an extended version of the LZ scenario.
The generalization of the basic LZ model takes place along two lines.
First, the dimensionality is enlarged to combine the description of
the external dynamics with the internal-state characterization. Second,
the model is extended to incorporate two avoided crossings emerging
from the changes induced in the energy bands by the modulation. Our
approach allows a first-principle derivation of the effective model-system
parameters. The obtained analytical results provide elements to control
the transitions. 
\end{abstract}
\maketitle

\section{Introduction}

The realization of Bose-Einstein condensates (BECs) with synthetic
spin-orbit coupling (SOC) \citep{key-SpielmanNature1} has opened
an active field of research where the progress in theory and experiments
is continuous \citep{key-ZhangSciRep}. The implementation of diverse
variations of the basic scenario have led to significant advances.
Particularly interesting are the extensions to different spin values
\citep{key-Spin1Spielman} or to fermionic systems \citep{key-fermionSOC1,key-fermionSOC2},
the study of excitations \citep{key-ZhangElemExciS1,key-zhangDipole,key-ZhangPhase,key-StringariSOC1,key-StringariSOC2,key-StringariSOCreview,key-finiteTempPhaseDiagram,key-collecmodeCenter}
with the analysis of dynamical instability \citep{key-F1SolitonInstSpielman,key-MithunModulInstability},
or the incorporation of optical lattices \citep{key-OpticalLattice}
and the research on superfluidity \citep{key-Superfluidity,key-SuperfluidityStringari}.
The variety of emerging physical effects, along with the possibility
of controlling the experimental conditions, allows regarding those
systems as a testing ground for the predictions of fundamental theories.
In this sense, it is worth mentioning the effort made in exploring
the potential appearance of novel states of matter, like nontrivial
superfluids or topological insulators \citep{key-SpielmanNature1,key-SpielmanRepProg,key-SpielmanToplolFeat,key-DalibardRMP}.
Crucial to the development of the field is the implementation of methods
to control the dynamics. Here, we focus on recent experiments, realized
in the \emph{standard} Raman-laser setup, which have uncovered the
possibility of implementing controlled variations of basic properties
of the system through the modulation of the Raman-coupling amplitude
\citep{key-OlsonStu}. Specifically, a sinusoidal modulation of the
coupling amplitude was found to induce nontrivial changes in the energy
bands and in the associated spin polarization. Moreover, gravitational
acceleration of the modulated system was shown to generate transitions
between the modified energy bands. An approach based on the application
of the Floquet formalism \citep{key-Anisimovas} was set up to explain
the findings. However, because of the complexity of the description,
no analytical insight was extracted from the complete Floquet picture.
In fact, it was in an operative reduction to a two-level system with
adjustable parameters where a satisfactory emulation of the experimental
results was produced. In particular, some general aspects of the transitions
were explained using an effective Landau-Zener (LZ) model \citep{key-Landau,key-Zener}
with two avoided crossings. The realization of Stueckelberg interferometry
\citep{key-NoriStuec} using the found structure of connected bands
served to test the theoretical framework: the achieved emulation of
the experimental output can be considered as a proof of consistency.
Although the compact character of the employed model allows envisaging
the control of the analyzed effects, the need of using adjustable
parameters seems to be a serious limiting factor. In our work, we
go further along the lines depicted in those preliminary analyses
by presenting \textcolor{black}{an approximate} analytical description
of the system response to the modulation. In our framework, set up
from first principles and with no adjustable parameters, the applicability
of a generalized LZ model will be rigorously shown. Our approach generalizes
the methodology used in previous work where related problems were
tackled. Namely, in \citep{key-SpielmanControl}, the use of a high-frequency
driving to modify in a controlled way the SOC magnitude, and, in turn,
the ground state, was experimentally evaluated. The reduced efficiency
of the method in particular spectral ranges was conjectured to be
due to heating. Identifying the heating mechanism was the objective
of the theoretical study of \citep{key-Llorente1}. There, a perturbative
scheme was found to provide the appropriate tools to explain the experimental
findings and to improve the performance of the proposed technique.
Additional related research was presented in \citep{key-OlsonGreene}.
There, experimental work on an unmodulated spin-orbit-coupled Bose-Einstein
condensate (SOBEC) uncovered the existence of inter-band transitions
induced by the acceleration of the system. In early analyses of those
experiments, an operative LZ model was applied to the characterization
of the transition probabilities \citep{key-OlsonGreene},\citep{key-XiongNonAdiabatic}.
Subsequently, in \citep{key-Llorente2}, an approximate analytical
description of the two, gravitational and harmonic, acceleration schemes
implemented in practice was provided. The scenario considered in those
previous studies must be generalized to deal with the system implemented
in \citep{key-OlsonStu}: first, it is necessary to evaluate the combined
effect of modulation and acceleration on a SOBEC; second, it is essential
going beyond the perturbative regime considered in \citep{key-Llorente1}.
Following those requirements, we present here a compact description
of the dynamics, where the mechanisms responsible for the observed
effects can be identified. Our approach is built up through a sequence
of unitary transformations. We start by transferring the modulation
of the Raman-coupling amplitude to multiple driving terms in the Hamiltonian.
By pondering the relative importance of those terms, we will be able
to single out the components, and, in turn, the mechanisms, that generate
the changes detected in the system properties. The procedure ends
with the derivation of a compact two-level Hamiltonian with a structure
similar to that of the model phenomenologically introduced in \citep{key-OlsonStu},
and, with effective parameters directly connected with the system
characteristics. In this approach, the observed energy bands are reproduced,
and, from the associated dressed states, the \emph{evolution} of the
spin polarization along the bands is analytically emulated. Additionally,
a generalized LZ model with two avoided crossings and the energy mismatch
linearly varying because of gravity is shown to be applicable to the
observed acceleration-induced transitions. Hence, our study supports
and generalizes the effective model employed in \citep{key-OlsonStu}.
The analytical character of the whole picture and the established
connection between the effective-model parameters and the system characteristics
provide us with some useful clues to implement strategies of control.

The outline of the paper is as follows. In Sec. II, the model system
and the basic elements of our methodology are introduced. In Sec.
III, we derive an effective undriven system which is shown to account
for the basic characteristics of the dynamics. Sec. IV is dedicated
to the description of the inter-band transitions induced by gravitational
acceleration. A generalized LZ model appropriate to describe those
transitions is derived and the effective parameters of the model are
identified. Finally, in Sec. V, a discussion of the implications and
potential extension of the study is presented. 

\section{Modulated spin-orbit coupling}

\subsection{The model system }

We focus on the realization of synthetic SOC in a one-dimensional
BEC reported in \citep{key-OlsonStu}. The coupling of the external
linear momentum of the atoms $\mathbf{p}$ to an effective ``spin'',
i.e., to a two-level internal system formed by hyperfine states, was
achieved through the use of two orthogonally polarized Raman lasers
with different propagation directions and frequencies. Specifically,
coupling dependent on the atom external momentum was implemented between
the three Zeeman-split states $\left|F,m_{F}\right\rangle $ of the
hyperfine $F=1$ ground-state manifold of $\textrm{R\ensuremath{b^{87}}}$.
Moreover, the detuning due to the quadratic Zeeman shift was used
to reduce the dynamics to the two levels associated to $m_{F}=0,-1$.
The corresponding states will be denoted as $\left|1,m_{F}=0\right\rangle \equiv\left|\downarrow\right\rangle $
and $\left|1,m_{F}=1\right\rangle \equiv\left|\uparrow\right\rangle $.
An essential component of the effective SOC created in \citep{key-OlsonStu}
was the modulation of the Raman-coupling amplitude $\Omega_{R}(t)$.
The characterization of the effects resulting from that modulation
constitutes the central objective of the present study. The system
is modeled by the (driven) Hamiltonian 

\begin{equation}
H_{SOC}(t)=\frac{P_{y}^{2}}{2m}+E_{r}\hat{1}+\frac{\hbar\Omega_{R}(t)}{2}\hat{\sigma}_{x}+\left(\frac{\hbar\delta}{2}+\frac{\alpha P_{y}}{\hbar}\right)\hat{\sigma}_{z},\label{eq:SOCHamiltonian}
\end{equation}
where $m$ denotes the atomic mass, $P_{y}$ stands for the momentum
operator, and $\sigma_{i}$ ($i=x,\,y\,,z$) are the Pauli matrices
corresponding to the \emph{pseudospin}, i.e., to the considered effective
two-level system. The rest of parameters refer to the laser characteristics:
$\alpha=2E_{r}/k_{r}$ is the strength of the realized SOC; $E_{r}=\hbar^{2}k_{r}^{2}/2m$
and $k_{r}=2\pi\sin(\theta/2)/\lambda$ are, respectively, the recoil
energy and momentum; $\lambda$ is the reference laser wavelength,
and, $\theta$ is the angle between the directions of the Raman lasers,
which, here, is fixed to $\theta=\pi/2$. Moreover, $\delta$ denotes
an adjustable detuning. In the considered experimental realization,
a sinusoidal modulation of the Raman-coupling amplitude $\Omega_{R}$,
with amplitude $\Omega_{M}$ and frequency $\omega_{M}$, was implemented.
Specifically, $\Omega_{R}$, with mean value $\Omega_{0}$, was varied
according to 

\begin{equation}
\Omega_{R}(t)=\Omega_{0}+\Omega_{M}\cos(\omega_{M}t).\label{eq:ModulationForm}
\end{equation}
 No harmonic confinement is considered. The relevance of this restriction
will be discussed further on. 

Before dealing with the effect of the modulation of the Raman amplitude,
it is pertinent to review some general characteristics of the undriven
scenario, i.e., of the system corresponding to take $\Omega_{M}=0$
in Eq. (\ref{eq:ModulationForm}). In former research \citep{key-SpielmanNature1,key-OlsonGreene},
general aspects of that system were analyzed using a single-particle
description as a first-stage approach. In particular, some properties
of the energy bands were described. From that work, one can extract
the following general considerations to take into account in our study:

First, since $P_{y}$ is a constant of the motion, it is appropriate
to use the basis $\left\{ \left|p_{y}\right\rangle \otimes\left|\chi\right\rangle \right\} $,
where $\left|p_{y}\right\rangle $ denotes a momentum state and $\left|\chi\right\rangle $
stands for a spin state. Then, the Hamiltonian can be reduced to the
spin space by simply replacing the operator $P_{y}$ by $\hbar q$
in Eq. (\ref{eq:SOCHamiltonian}). ($q$ is the quasi-momentum in
the coupling direction, $p_{y}=\hbar q$). Hence, it is straightforwardly
shown that the system presents the lower and upper (undriven) energy
bands, $E_{-}^{(ud)}$ and $E_{+}^{(ud)}$, given by 

\begin{equation}
E_{\pm}^{(ud)}(q)=\frac{\hbar^{2}q^{2}}{2m}+E_{r}\pm\sqrt{\left(\frac{\hbar\Omega_{0}}{2}\right)^{2}+\left(\frac{\hbar\delta}{2}+\alpha q\right)^{2}}.\label{eq:bands}
\end{equation}

Moreover, since the bands correspond to eigenvalues of the reduced
Hamiltonian, no transitions between them are predicted to occur in
this scheme. Indeed, the bands reflect the existence of locking between
the momentum and the spin state: each quasi-momentum value $q$ is
attached to a particular pair of eigenvalues, $E_{-}^{(ud)}(q)$ and
$E_{+}^{(ud)}(q)$, and, in turn, to the corresponding pair of internal
states. 

Interband transitions induced by the acceleration of the system, i.e.,
by the variation of the quasi-momentum, were studied in this (undriven)
setting. Indeed, acceleration resulting from the effect of gravity
was found to generate transitions which were satisfactorily described
using the LZ model. The \emph{activation }of gravity was arranged
in practice by implementing the SOC in the vertical axis. Then, the
trap was turned off, and the system was led to evolve under the effect
of gravity. A qualitative understanding of the mechanism responsible
for those interband transitions can be achieved from the analysis
of Fig. 1(a) where the bands given by Eq. (\ref{eq:bands}) are depicted.
The curves $E_{\pm}^{(ud)}(q)$ can be regarded as corresponding to
the adiabatic levels (eigenvalues obtained for a \emph{frozen} quasimomentum
$q$) of an effective LZ system. \textcolor{black}{In this picture,
the point of maximum proximity between the curves can be identified
with the avoided crossing characteristic of the LZ model}. In this
framework, the inclusion of the acceleration, i.e., of varying $q$,
implies going beyond that picture to take into account the potential
occurrence of nonadiabatic effects, i.e., of interband transitions.
The present work will be focused on analyzing how this scenario of
LZ transitions is modified by the modulation of the Raman-coupling
amplitude. Given the weakly-harmonic confinement realized in practice,
a quasi-continuum approximation is feasible, and the use of the term
``band'' is still appropriate. When stronger harmonic confinement
is considered, the description becomes more complex since the eigenstates
of the Hamiltonian do not have a well-defined momentum (they are not
eigenstates of $P_{y}$). Actually, since the harmonic trapping alters
the evolution of the momentum states, it can be regarded as an acceleration
mechanism alternative to that associated to gravity. 

\subsection{Transferring the modulation to multiple driving terms }

The introduction of the modulation implies dealing with a time-dependent
Hamiltonian. In our procedure to describe the dynamics, we start by
applying the unitary transformation

\begin{equation}
U_{1}(t)=\exp\left[-\frac{i}{2}\zeta(t)\hat{\sigma}_{x}\right],\label{eq:3}
\end{equation}
 where we have defined

\begin{equation}
\zeta(t)\equiv\eta_{M}\sin(\omega_{M}t),
\end{equation}
with $\eta_{M}=\Omega_{M}/\omega_{M}$. The transformed Hamiltonian,
obtained from the expression $U_{1}^{\dagger}H_{SOC}U_{1}-i\hbar U_{1}^{\dagger}\dot{U}_{1}$,
is written, after straightforward algebra, and, employing the same
notation as that used for the untransformed Hamiltonian, in the form 

\textcolor{black}{
\begin{eqnarray}
H_{SOC}(t) & = & \left(\frac{\hbar^{2}q^{2}}{2m}+E_{r}\right)\hat{1}+\frac{\hbar\Omega_{0}}{2}\hat{\sigma}_{x}+\left(\frac{\hbar\delta}{2}+\alpha q\right)\left[\hat{\sigma}_{z}\cos\zeta(t)+\hat{\sigma}_{y}\sin\zeta(t)\right].\label{eq:HamiltSOCBECFirstTrans}
\end{eqnarray}
} Emulating the experimental setup \citep{key-OlsonStu}, where the
detuning was adjusted to zero, we will take $\delta=0$. The implications
of a nonzero detuning will be discussed further on. Now, employing
the expansion of $\cos\zeta(t)$ and $\sin\zeta(t)$ in terms of the
ordinary Bessel functions $J_{n}(\eta_{M})$ \citep{key-Grad}, $H_{SOC}(t)$
is rewritten as the sum of a basic undriven Hamiltonian and different
oscillating contributions, namely,  

\begin{eqnarray}
H_{SOC}(t) & = & \left(\frac{\hbar^{2}q^{2}}{2m}+E_{r}\right)\hat{1}+\frac{\hbar\Omega_{0}}{2}\hat{\sigma}_{x}+J_{0}(\eta_{M})\alpha q\hat{\sigma}_{z}+\nonumber \\
 &  & 2J_{1}(\eta_{M})\alpha q\hat{\sigma}_{y}\sin(\omega_{M}t)+\sum_{n>1}2J_{n}(\eta_{M})\alpha q\hat{\sigma}_{i}^{(n)}\cos(n\omega_{M}t),\label{eq:DevelopDrivenH}
\end{eqnarray}
 where $\hat{\sigma}_{i}^{(n)}=\hat{\sigma}_{z}$ when $n$ is even,
and $\hat{\sigma}_{i}^{(n)}=\hat{\sigma}_{y}$ when $n$ is odd. 

The above expansion of the Hamiltonian was used in \citep{key-Llorente1}
to describe the dynamics in a former scenario of Raman-amplitude modulation
\citep{key-SpielmanControl}. Specifically, Eq. (\ref{eq:DevelopDrivenH})
was applied  to set up a perturbative scheme that allowed the design
of strategies to control the dynamics by varying the driving parameters.
It is worth noticing that the strength of the modified zero-order
term can be altered by varying the factor $J_{0}(\eta_{M})$ and the
amplitude of the different oscillating contributions can be controlled
by modifying $J_{1}(\eta_{M})$ and $J_{n}(\eta_{M})$ ($n>1$). The
presence of those factors in reduced descriptions of systems with
sinusoidal driving has been frequently used to design methods of control
in different contexts \citep{key-Llorente3}. The experimental results
of \citep{key-SpielmanControl} were satisfactorily explained using
a model based on keeping the first driving term and neglecting, via
a coarse-graining, the higher-order (oscillating) components (i.e.,
those with $n>1$) \citep{key-Llorente1}. Actually, the applicability
of the averaging method is justified for small modulation amplitudes
and/or high frequencies: as $\eta_{M}=\Omega_{M}/\omega_{M}$ decreases,
the functions $J_{n}(\eta_{M})$ ($n\geq1$) decay and eventually
go to zero. In the present work, we will keep on neglecting the contribution
of those higher-order terms. However, in the analysis of the first
oscillating component, we will go beyond the perturbative scheme used
in \citep{key-Llorente1}. 

\section{The effect of the modulation on the spin-momentum bands }

\subsection{Setting up an effective undriven framework}

In \citep{key-OlsonStu}, the dynamics of the modulated system was
described applying the Floquet formalism. However, in order to operatively
account for the experimental results, a functional \emph{ad hoc} simplification
of the Floquet framework was implemented: an effective time-independent
two-level system with parameters adjusted to reproduce the experimental
findings was employed. Here, we will apply an (alternative) approach
based on a sequence of unitary transformations which will allow us
to trace the applicability of an effective time-independent Hamiltonian
from first principles. In order to simplify our procedure, we first
rewrite the Hamiltonian given by Eq. (\ref{eq:DevelopDrivenH}), leaving
out the terms with $n>1$, as 
\begin{equation}
H_{SOC}(t)=a(q)\hat{1}+b\hat{\sigma}_{x}+c(q)\hat{\sigma}_{z}+d(q)\hat{\sigma}_{y}\sin(\omega_{M}t),\label{eq:CompactSOCHamiltonian}
\end{equation}
 where,

\begin{eqnarray}
a(q) & = & \frac{\hbar^{2}q^{2}}{2m}+E_{r},\\
b & = & \frac{\hbar\Omega_{0}}{2},\\
c(q) & = & J_{0}(\eta_{M})\alpha q,\\
d(q) & = & 2J_{1}(\eta_{M})\alpha q.
\end{eqnarray}
 Moreover, using the function $\beta(q)$ defined by 

\begin{equation}
\beta(q)=\arctan\left[\frac{b}{c(q)}\right],
\end{equation}
Eq. (\ref{eq:CompactSOCHamiltonian}) is recast as 
\begin{equation}
H_{SOC}(t)=a(q)\hat{1}+\sqrt{b^{2}+c(q)^{2}}\left[\hat{\sigma}_{x}\sin\beta(q)+\hat{\sigma}_{z}\cos\beta(q)\right]+d(q)\sin(\omega_{M}t)\hat{\sigma}_{y},
\end{equation}
 Now, rotating an angle $\beta(q)$ around the $OY$ axis via the
unitary transformation 

\begin{equation}
U_{2}=e^{-i\beta(q)\sigma_{y}/2},
\end{equation}
the Hamiltonian is converted into

\begin{equation}
U_{2}^{\dagger}H_{SOC}U_{2}=a(q)\hat{1}+\sqrt{b^{2}+c(q)^{2}}\hat{\sigma}_{z}+d(q)\sin(\omega_{M}t)\hat{\sigma}_{y}\equiv H_{SOC}(t).
\end{equation}
 Furthermore, in the rotating frame defined by 

\begin{equation}
U_{3}(t)=e^{-i\omega_{M}t\sigma_{z}/2},
\end{equation}
 one has 

\begin{eqnarray}
U_{3}^{\dagger}H_{SOC}U_{3}-i\hbar U_{3}^{\dagger}\dot{U}_{3} & = & a(q)\hat{1}+\left(\sqrt{b^{2}+c(q)^{2}}-\hbar\omega_{M}/2\right)\hat{\sigma}_{z}+\nonumber \\
 &  & d(q)\sin(\omega_{M}t)\left[\cos(\omega_{M}t)\hat{\sigma}_{y}+\sin(\omega_{M}t)\hat{\sigma}_{x}\right]\equiv H_{SOC}(t).
\end{eqnarray}
 Finally, making the Rotating Wave Approximation (RWA) \citep{key-CohenBookAtomPhoton},
i.e., keeping the secular terms and neglecting oscillations of frequency
$2\omega_{M}$, we arrive at 

\begin{equation}
H_{SOC}=a(q)\hat{1}+e(q)\hat{\sigma}_{z}+\frac{d(q)}{2}\hat{\sigma}_{x},\label{eq:EffUndrivHamilt}
\end{equation}
 with

\begin{equation}
e(q)=\sqrt{b^{2}+c(q)^{2}}-\hbar\omega_{M}/2.\label{eq:e(q)}
\end{equation}
 Notice that the restriction $\left|d(q)\right|\ll\omega_{M}$ must
be fulfilled to guarantee the validity of the RWA. Observe also that
the RWA is consistent with neglecting the higher-order oscillating
terms in Eq. (\ref{eq:DevelopDrivenH}).

Hence, we have reached an effective time-independent Hamiltonian where
the modulation of the Raman coupling amplitude has been included into
the effective \emph{parameters} $e(q)$ and $d(q)$. Specifically,
the splitting term $e(q)$ incorporates the shift $\hbar\omega_{M}/2$,
and, through $c(q)$, the driving-dependent factor $J_{0}(\eta_{M})$;
furthermore, the coupling term $\frac{d(q)}{2}$ absorbs the factor
$J_{1}(\eta_{M})$. Observe that both \emph{parameters} $e(q)$ and
$d(q)$ depend on the quasimomentum, i.e., the SOC enters two terms
of the effective Hamiltonian. In the following, this framework will
be shown to provide a valuable description of the system dynamics.

\subsection{Dressed states and modified spin-momentum locking}

The eigenvalues of the Hamiltonian given by Eq. (\ref{eq:EffUndrivHamilt})
are readily obtained: 
\begin{equation}
E_{\pm}(q)=a(q)\pm\left[e^{2}(q)+\frac{d^{2}(q)}{4}\right]^{1/2}.\label{eq:NewBands}
\end{equation}
 Additionally, going back in the sequence of unitary transformations,
the associated (dressed) eigenstates $\left\{ \left|\chi_{\pm}\right\rangle \right\} $
are expressed in the original basis $\left\{ \left|\downarrow\right\rangle ,\left|\uparrow\right\rangle \right\} $
as

\begin{eqnarray*}
\left|\chi_{\pm}\right\rangle  & = & F_{\pm,\uparrow}(t)\left|\uparrow\right\rangle +F_{\pm,\downarrow}(t)\left|\downarrow\right\rangle 
\end{eqnarray*}
 where

\begin{eqnarray}
F_{\pm,\uparrow}(t) & = & C_{\pm,1}e^{i\omega_{M}t}(\cos\frac{\beta}{2}\cos\frac{\zeta}{2}-i\sin\frac{\beta}{2}\sin\frac{\zeta}{2})+\nonumber \\
 &  & C_{\pm,2}(\sin\frac{\beta}{2}\cos\frac{\zeta}{2}+i\cos\frac{\beta}{2}\sin\frac{\zeta}{2}),
\end{eqnarray}
 and

\begin{eqnarray}
F_{\pm,\downarrow}(t) & = & C_{\pm,1}e^{i\omega_{M}t}(-\sin\frac{\beta}{2}\cos\frac{\zeta}{2}+i\cos\frac{\beta}{2}\sin\frac{\zeta}{2})+\\
 &  & C_{\pm,2}(\cos\frac{\beta}{2}\cos\frac{\zeta}{2}+i\sin\frac{\beta}{2}\sin\frac{\zeta}{2}),
\end{eqnarray}
with

\begin{equation}
C_{\pm,1}=\frac{d(q)/2}{\left(\frac{d^{2}(q)}{4}+\left[a(q)+e(q)-E_{\pm}\right]^{2}\right)^{1/2}},
\end{equation}
 and

\begin{equation}
C_{\pm,2}=-\frac{a(q)+e(q)-E_{\pm}}{\left(\frac{d^{2}(q)}{4}+\left[a(q)+e(q)-E_{\pm}\right]^{2}\right)^{1/2}},
\end{equation}
In turn, the spin polarization in each of the bands is straightforwardly
calculated, namely,

\begin{equation}
\left\langle \chi_{\pm}\right|\hat{\sigma}_{z}\left|\chi_{\pm}\right\rangle =\left|F_{\pm,\uparrow}(t)\right|^{2}-\left|F_{\pm,\downarrow}(t)\right|^{2}.
\end{equation}
 Moreover, the averaging over the driving period $T_{M}=2\pi/\omega_{M}$
is directly carried out. In particular, for the averaged polarization
along the lower band, one finds 

\begin{eqnarray}
\left\langle \left\langle \chi_{-}\right|\hat{\sigma}_{z}\left|\chi_{-}\right\rangle \right\rangle _{T_{M}} & = & \left(\left|C_{-,1}\right|^{2}-\left|C_{-,2}\right|^{2}\right)\cos\beta\left[J_{0}^{2}(\eta_{M}/2)-2J_{1}^{2}(\eta_{M}/2)\right]+\nonumber \\
 &  & 4C_{-,1}C_{-,2}J_{0}(\eta_{M}/2)J_{1}(\eta_{M}/2).
\end{eqnarray}

From these results, some preliminary conclusions can be drawn: 

i) The band structure is significantly altered by the modulation.
The eigenvalues $E_{\pm}(q)$, depicted in Fig. 1(b), present significant
differences with the undriven bands, contained in Fig. 1(a). Particularly
conspicuous is the emergence of two-avoided crossings in the modulated
system. This feature is specifically rooted in the driving: from Eq.
(\ref{eq:bands}), the difference between the undriven eigenvalues
$E_{+}^{(ud)}$ and $E_{-}^{(ud)}$ is found to present a single minimum,
reached at $q=0$; in contrast, in the modulated case, the difference
between the energies $E_{\pm}(q)$, given by Eq. (\ref{eq:NewBands}),
display two minima, essentially because of the shift $\hbar\omega_{M}/2$
\textcolor{black}{in Eq. (\ref{eq:e(q)})}. Importantly, it is apparent
that the separation between the avoided crossings increases with $\omega_{M}$.
In the next section, the relevance of this property to extending the
LZ model to deal with two avoided crossings will be apparent. \bigskip{}

\includegraphics{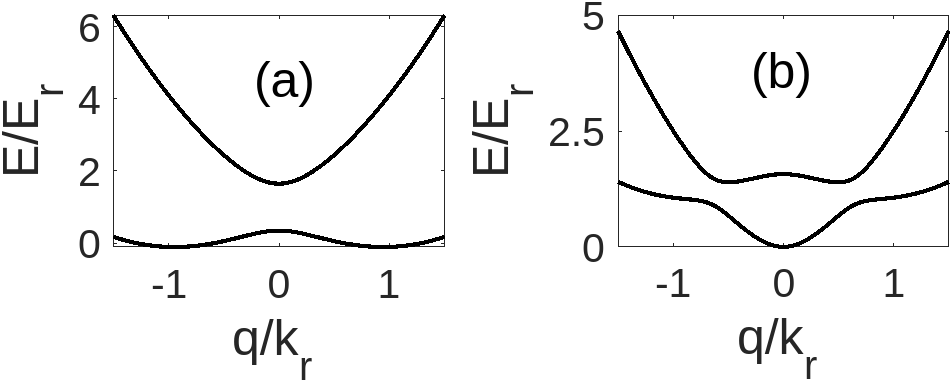}

\begin{figure}[H]

\caption{Energy-momentum dispersion relations for the unmodulated system (a)
and for the modulated system (b) {[}In both cases, $\Omega_{0}=1.3E_{r}/\hbar$,
$\delta=0$; additionally, $\Omega_{M}=1.3E_{r}/\hbar$ and $\omega_{M}=2.87E_{r}/\hbar$
for (b){]}. (In each case, the energy origin has been shifted to the
lowest point in the first band).}

\end{figure}

ii) Additional information is extracted from the analysis of the dressed
states. As can be seen in Fig. 2, where results for the polarization
along the lower band are presented for the unmodulated state {[}Fig.
2(a){]} and for the modulated (dressed) state {[}Fig. 2(b){]}, the
driving of the Raman coupling amplitude alters significantly the distribution
of population in the states $\left\{ \left|\downarrow\right\rangle ,\left|\uparrow\right\rangle \right\} $.
The monotonous character of the evolution of the polarization in the
undriven case disappears when the modulation is applied. The analytical
character of our results allows choosing the modulation parameters
to alter the polarization pattern. The experimental results presented
in \citep{key-OlsonStu} are reproduced by our findings. No adjustable
parameters have been employed. \textcolor{black}{Observe that, whereas,
in Fig. 2d in \citep{key-OlsonStu}, the experimental results for
the spin polarization are displayed as a function of time, our findings
are presented as a function of the normalized quasimomentum.}\bigskip{}

\includegraphics{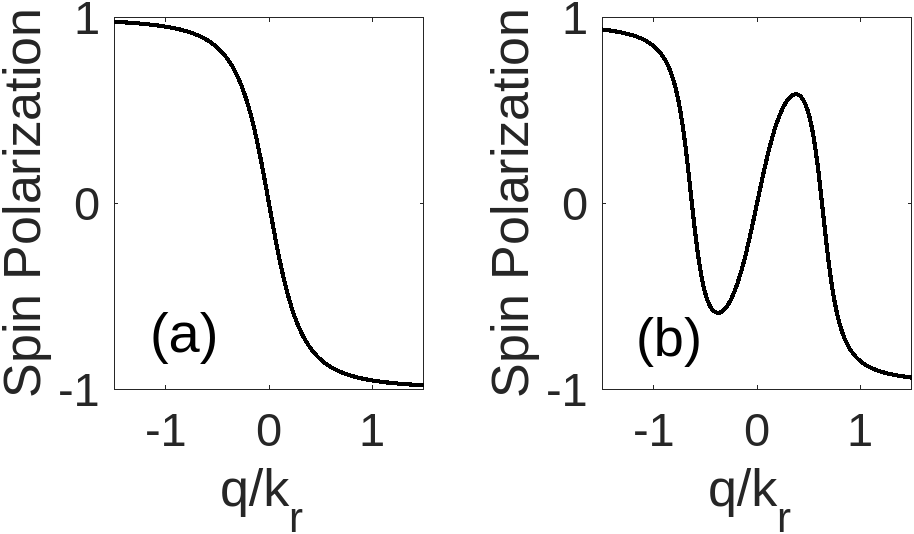}

\begin{figure}[H]

\caption{The spin polarization vs. the quasimomentum for the unmodulated system
(a), and, for the modulated system (b). (Same parameters as in Fig.
1).}

\end{figure}

iii) Our results improve the effective two-level system operatively
introduced in \citep{key-OlsonStu}. Two main contributions of our
study must be remarked. First, our approach traces the presence of
the driving-dependent factor $J_{0}(\eta_{M})$ in the splitting term
$e(q)$ of the model. Second, from our results, the effective coupling
term is identified as $\frac{d(q)}{2}=J_{1}(\frac{\Omega_{M}}{\omega_{M}})\alpha q$.
Here, it is worth recalling that the operative model of \citep{key-OlsonStu}
incorporates an adjustable effective coupling parameter, which was
found to display a linear dependence on the modulation amplitude $\varOmega_{M}$.
That finding is confirmed and generalized by our study: the actual
dependence of the effective coupling on $\varOmega_{M}$ is that included
into $J_{1}(\frac{\Omega_{M}}{\omega_{M}})$. The linear dependence
is specific to the range of amplitudes and modulation frequency considered
in the experiment. Indeed, taking into account the approximation \citep{key-Grad}

\begin{equation}
J_{1}(\frac{\Omega_{M}}{\omega_{M}})\simeq\frac{1}{2}\frac{\Omega_{M}}{\omega_{M}}+\mathcal{O}\left[\left(\frac{\Omega_{M}}{\omega_{M}}\right)^{3}\right],
\end{equation}
 one accounts for the linear behavior found in \citep{key-OlsonStu}.
This agreement is illustrated in \textcolor{black}{Fig. 3}, where
the coupling parameter emergent in our model is represented as a function
of $\varOmega_{M}$. Notice that the range of $\Omega_{M}/\omega_{M}$
where the decline of $J_{1}(\frac{\Omega_{M}}{\omega_{M}})$ takes
place is outside the regime explored in the experiments. Our results
open the possibility of controlling the coupling by varying either
$\varOmega_{M}$ or $\omega_{M}$.\bigskip{}

\includegraphics[scale=0.8]{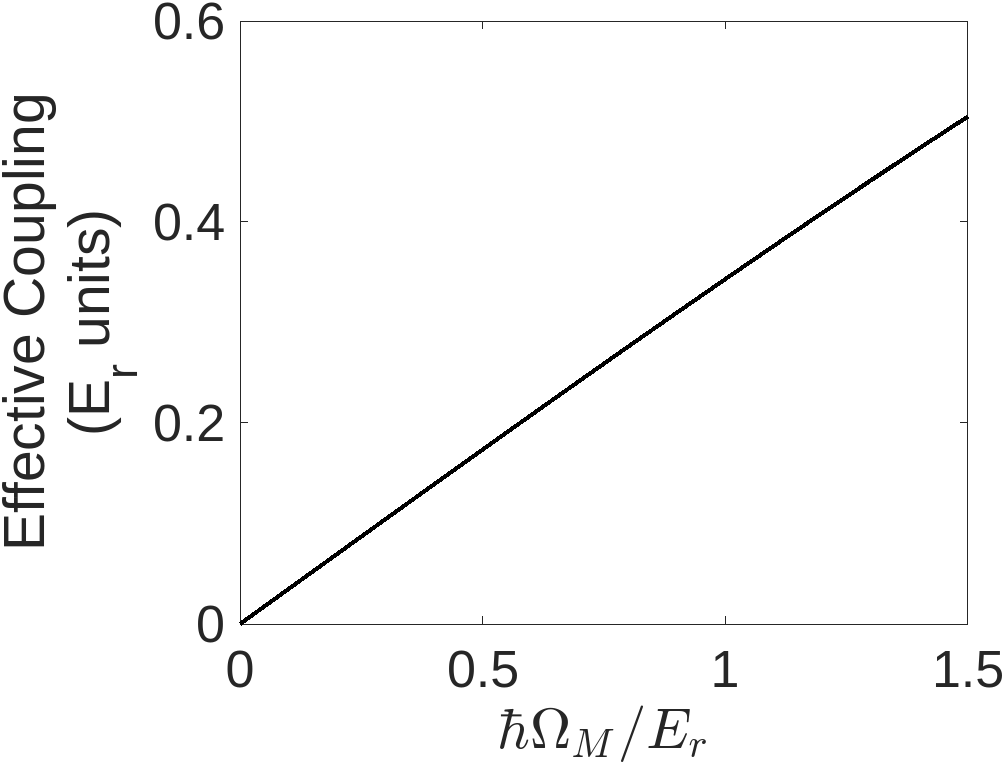}

\begin{figure}[H]

\caption{The effective-coupling term $\frac{d(q)}{2}=J_{1}(\frac{\Omega_{M}}{\omega_{M}})\alpha q$
as a function of the modulation amplitude. ($\Omega_{0}=1.3E_{r}/\hbar$,
$\delta=0$, $\omega_{M}=2.87E_{r}/\hbar$, $q=1$ ). }

\end{figure}

iv) An additional feature uncovered by our results is the linear dependence
of the effective coupling term on both, the SOC strength $\alpha$
and the quasimomentum $q$. In the next section, we will see the relevance
of this finding to the understanding of the mechanism responsible
for acceleration-induced transitions.

\section{Landau-Zener transitions induced by gravitational acceleration}

A significant achievement of the experiments of \citep{key-SpielmanControl}
was the uncovering of interband transitions induced by gravitational
acceleration. In the practical arrangement, after implementing the
SOC in the vertical axis, the trap was switched off and the system
evolution under the effect of gravity was monitored. The changes in
the band populations were tentatively explained using a LZ model with
optimized characteristic parameters. Here, we will rigorously trace
the emergence of that model. 

According to the practical implementation, we consider that, at $t=0$,
the gravitational field is \emph{activated}, the dynamics being then
governed by the Hamiltonian 
\begin{eqnarray}
H_{G} & = & H_{SOC}+H_{grav}\nonumber \\
 & = & \frac{P_{y}^{2}}{2m}+E_{r}\hat{1}+\frac{\hbar\Omega_{R}(t)}{2}\hat{\sigma}_{x}+\frac{\alpha P_{y}}{\hbar}\hat{\sigma}_{z}+mgY,\label{eq:HamiltG}
\end{eqnarray}
where, for $H_{SOC}$ we have used the expression given by Eq. (\ref{eq:SOCHamiltonian})
with $\delta=0$, and, the effect of the gravitational field has been
assumed to be well modeled by the term $H_{grav}=mgY$. As shown in
\citep{key-kasevich}, this approach is convenient to operatively
deal with the effects of gravity in the present context. 

In the setup provided by Eq. (\ref{eq:HamiltG}), $P_{y}$ is not
a constant of the motion. The incorporation of gravity implies that
the eigenstates of the Hamiltonian have no longer the form $\left|p_{y}\right\rangle \otimes\left|\chi\right\rangle $.
In other terms, $H_{grav}$ leads to the variation of the momentum
in the evolution of a state $\left|p_{y}\right\rangle \otimes\left|\chi\right\rangle $,
i.e., it induces the acceleration of the system. 

Our procedure to deal with the dynamics resulting from the \emph{incorporation}
of gravity includes the consecutive application of two unitary transformation.
The first one, given by 

\begin{equation}
U_{4}(t)=\exp\left[-\frac{i}{\hbar}mgYt\right],\label{eq:Udesplaz}
\end{equation}
 introduces a time-dependent displacement in momentum: it implies
working in a reference frame translated with acceleration $g$. The
transformed Hamiltonian reads 

\begin{eqnarray*}
U_{4}^{\dagger}H_{G}U_{4}-i\hbar U_{4}^{\dagger}\dot{U}_{4} & = & \frac{(P_{y}-mgt)^{2}}{2m}+E_{r}\hat{1}+\frac{\hbar\Omega_{R}(t)}{2}\hat{\sigma}_{x}+\frac{\alpha(P_{y}-mgt)}{\hbar}\hat{\sigma}_{z}\equiv H_{G}.
\end{eqnarray*}
Further simplification is achieved by applying the transformation

\begin{equation}
U_{5}(t)=\exp\left[-\frac{i}{\hbar}\left(\frac{P_{y}^{2}}{2m}+E_{r}\hat{1}\right)t+\frac{i}{\hbar}gP_{y}\frac{t^{2}}{2}\right],\label{eq:Ugauge-1}
\end{equation}
which leads us to write the Hamiltonian in the form 
\begin{equation}
U_{5}^{\dagger}H_{G}U_{5}-i\hbar U_{5}^{\dagger}\dot{U}_{5}=\frac{\hbar\Omega_{R}(t)}{2}\hat{\sigma}_{x}+\frac{\alpha(P_{y}-mgt)}{\hbar}\hat{\sigma}_{z}\equiv H_{G},
\end{equation}
 where we have left out the dynamically irrelevant term $\frac{1}{2}mg^{2}t^{2}$,
which simply shifts the energy origin by $mg^{2}t^{2}/2$. Now, working
in the representation $\left\{ \left|p_{y}\right\rangle \otimes\left|\chi\right\rangle \right\} $,
the reduction to the spin space is readily implemented in the form

\begin{equation}
H_{G}=\frac{\hbar\Omega_{R}(t)}{2}\hat{\sigma}_{x}+\alpha q_{g}\hat{\sigma}_{z},
\end{equation}
 with the \emph{shifted} quasimomentum 

\begin{equation}
q_{g}=\frac{p_{y}}{\hbar}-\frac{mg}{\hbar}t\equiv q-\frac{mg}{\hbar}t.\label{eq:displQuasiM}
\end{equation}
 Finally, the application of a sequence of unitary transformations
similar to those introduced in the previous section, i.e., the consecutive
application of $U_{1}$, $U_{2}$, and, $U_{3}$, leads to 

\begin{equation}
H_{G}=e(q_{g})\hat{\sigma}_{z}+\frac{d(q_{g})}{2}\hat{\sigma}_{x}.\label{eq:FinalEffectLZ}
\end{equation}
Here, it is worth pointing out that additional terms emerge when applying
the transformation $U_{2}(t)=e^{-i\beta(q_{g})\sigma_{y}/2},$ which
is now time dependent because of the shift incorporated in $q_{g}$.
Those terms are rooted in elements of the undriven dynamics as can
be seen from the definition of $\beta(q_{g})$. They are indeed relevant
to recover the results obtained for the undriven scenario in \citep{key-Llorente2}:
they account for LZ transitions in the absence of driving. However,
in the case considered here, where the modulation induces a significant
modification of the energy bands, those terms can be considered to
give a (nonsecular) correction to $H_{G}$. Hence, we keep them out
of the description. 

Let us see that it is possible to identify LZ characteristics in the
dynamics resulting from Eq. (\ref{eq:FinalEffectLZ}). To this end,
it is convenient to recall that the basic LZ model corresponds to
a two-level system described by the Hamiltonian 
\begin{equation}
H_{LZ}=\hbar\frac{vt}{2}\hat{\sigma}_{z}+\hbar\zeta\hat{\sigma}_{x},\label{eq:BasicLZmodel}
\end{equation}
where $v$ stands for the variation rate of the energy mismatch, and
$\zeta$ denotes the interstate coupling strength. Whereas the diabatic
states are the two (coupled) eigenstates of $\sigma_{z}$, the adiabatic
states are the instantaneous eigenstates of the whole Hamiltonian
$H_{LZ}$. At $t=0,$ there is a crossing of the diabatic levels which
becomes an avoided crossing in the adiabatic picture. 

From the comparison of the basic LZ scenario with our effective description
of the modulated system, the following conclusions can be drawn:

i) By taking \emph{frozen} values of the quasimomentum $q_{g}$, the
adiabatic levels of the Hamiltonian given by Eq. (\ref{eq:FinalEffectLZ})
are readily obtained. They are depicted in Fig. 4(b). Notice that
they are actually related to the dispersion curves (\emph{dressed
}bands) given by Eq. (\ref{eq:NewBands}) and represented in Fig.
1(b). The differences are rooted in the terms removed via the application
of the unitary transformation $U_{5}(t)$ in the derivation of $H_{G}$.
The acceleration leads to changes in the quasimomentum $q_{g}$, and,
consequently, implies going beyond the adiabatic picture, accounting
then for the occurrence of inter-band transitions. 

ii) In the diabatic picture, the energy levels are the \emph{bare
}bands obtained by taking out the contribution of the off-diagonal
term $\frac{d(q_{g})}{2}\hat{\sigma}_{x}$ in Eq. (\ref{eq:FinalEffectLZ}).
Correspondingly, that term is now regarded as giving the coupling
between the (diabatic) spin states. The diabatic levels are displayed
in Fig. 4(a). This picture is appropriate to illustrate the identification
of one of the characteristic parameters of the LZ model: the variation
rate of the energy mismatch is obtained from the linearization of
the diabatic levels at the crossings. One must stress that, in the
primary LZ setup, it is the driving term, $\hbar\frac{vt}{2}\sigma_{z}$,
independent of the considered dynamics, that induces the linear variation
of the energy mismatch, and, eventually, the crossing. In contrast,
in our system, it is the dynamics of the external variables, specifically,
of the momentum, that leads to the modification of the internal-state
splitting. In the applied reduction to the spin space, that dynamical
effect takes the form of a driving component, namely, it turns into
the form $e(q-\frac{mg}{\hbar}t)\hat{\sigma}_{z}$.\bigskip{}

\includegraphics{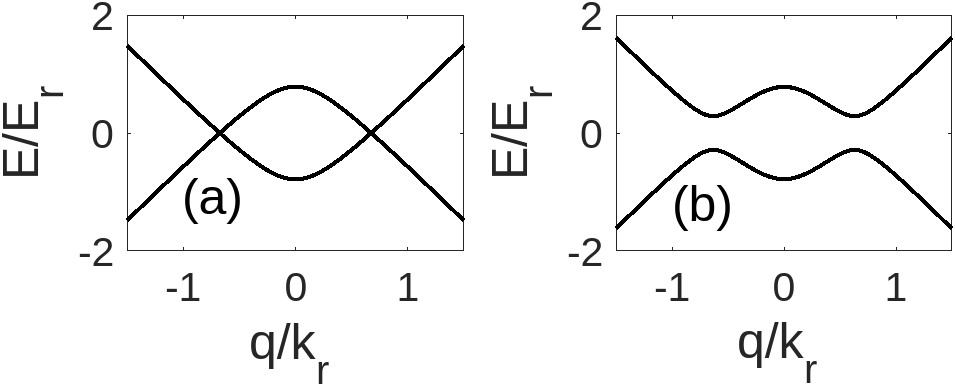}

\begin{figure}[H]

\caption{Diabatic (a) and adiabatic (b) energy-levels vs the quasimomentum
for the effective LZ model. (Same parameters as for the modulated
system in Fig. 1). }

\end{figure}

iii) An important differential feature of the present scenario is
the existence of two avoided crossings, instead of just one as in
the basic LZ model. Since the crossings are defined by the condition
$e(q_{C})=0$, it follows from Eq. (\ref{eq:e(q)}) that they are
reached at the quasimomentum values

\begin{equation}
q_{C}=\pm\frac{\hbar}{2J_{0}(\eta_{M})\alpha}\sqrt{\omega_{M}^{2}-\Omega_{0}^{2}}.\label{eq:CrossingQ}
\end{equation}
 The associated gaps in the adiabatic curves are determined by the
effective\emph{ offset} $d(q_{C})$: the adiabatic bands progressively
separate as $d(q_{C})$ increases. Note that the condition $\omega_{M}>\Omega_{0}$
must be fulfilled for a crossing to take place. Otherwise, the parallelism
with the LZ model cannot be established.

iv) The identification of the effective parameters $v_{eff}$ and
$\zeta_{eff}$ of the modulated system, (i.e., the counterparts of
$v$ and $\zeta$ of the basic LZ model), is straightforward. In order
to identify $v_{eff}$, the linearization of the energy mismatch at
the level crossings is required. Specifically, one obtains that, near
the crossing points, the time dependence of the energy splitting can
be approximated as 

\begin{equation}
e(t)\sim-\alpha\frac{mg}{\hbar}\frac{\sqrt{\omega_{M}^{2}-\Omega_{0}^{2}}}{\omega_{M}}J_{0}(\eta_{M})(t-t_{C}),\label{eq:approximateDetuning}
\end{equation}
 where $t_{C}$, which denotes the time at the crossing, depends on
the initial momentum preparation. We do not need to specify the value
taken by $t_{C}$: it is well-known from the study of the LZ model
\citep{key-Llorente2} that the existence of an independent term in
the linearly-modified energy mismatch does not affect the probability
of transition 

From the obtained rate of variation of the energy splitting at the
crossing, one can establish the correspondence 

\begin{equation}
\left|v_{eff}\right|\rightarrow\frac{2\alpha mg}{\hbar^{2}}\frac{\sqrt{\omega_{M}^{2}-\Omega_{0}^{2}}}{\omega_{M}}J_{0}(\eta_{M}).\label{eq:EffLZparam1}
\end{equation}
 Apart from the \emph{expected} dependence on $g$, $v_{eff}$ shows
the relevance of the SOC strength $\alpha$ and of the modulation
frequency $\omega_{M}$ to the rate of variation of the energy mismatch. 

Additionally, the effective parameter $\zeta_{eff}$ is traced from
the value of the coupling term at the crossing: 

\begin{equation}
\zeta_{eff}\rightarrow\frac{1}{\hbar}\frac{d(q_{C})}{2}=\frac{J_{1}(\eta_{M})}{2J_{0}(\eta_{M})}\sqrt{\omega_{M}^{2}-\Omega_{0}^{2}}.\label{eq:EffLZparam2}
\end{equation}
 It is pertinent to recall that the applied framework contains elements
which are intrinsically associated to the modulation characteristics.
Namely, the approach incorporates the condition $\omega_{M}>\Omega_{0}$,
imposed to establish the parallelism with the LZ model. Additionally,
it makes use of a reference frame rotating with $\omega_{M}$. Moreover,
corrections associated to the transitions already present in the undriven
system have been left out. Consequently, the present framework cannot
account for the recovery of the results obtained for the unmodulated
system \citep{key-Llorente2}.

v) In order to guarantee the applicability of the LZ expressions,
the initial and final times of the process do not need to strictly
correspond to $t\rightarrow\mp\infty$. The robust character of the
LZ predictions imply that it is only necessary to work with sufficiently
large values of the initial and final splittings. Observe that the
separation between crossing points increase with the modulation frequency.
Then, for sufficiently large values of $\omega_{M}$, one can assume
that the \emph{asymptotic} regime is reached between the crossings,
and, consequently, that the LZ model can be safely applied to each
of them. This is in fact the basis of Stueckelberg interferometry
\citep{key-NoriStuec}. \textcolor{black}{Let us make a more precise
evaluation of the required magnitude of $\omega_{M}$. It is apparent
that a sufficient condition for effectively reaching the asymptotic
regime between crossings is to impose the duration of the LZ transition
to be much smaller than the time taken for the parameter $q_{g}(t)$
to go from one crossing to the other. This condition can be more precisely
stated in the following form: }

\textcolor{black}{From basic work on the LZ model, it is known that
the typical duration of the LZ transition $\tau_{LZ}$, i.e., the
interval required for reaching the asymptotic population values, can
be estimated as 
\[
\tau_{LZ}\sim\frac{\zeta_{eff}}{\left|v_{eff}\right|}.
\]
 Additionally, from our study, the quasimomentum separation between
the two avoided crossings is shown to be given by }

\textcolor{black}{
\[
\frac{\hbar}{J_{0}(\eta_{M})\alpha}\sqrt{\omega_{M}^{2}-\Omega_{0}^{2}}.
\]
 In turn, the time $\tau_{CC}$ taken for the variation of the parameter
$q_{g}(t)$ between crossings is given by }

\textcolor{black}{
\[
\tau_{CC}\sim\frac{\frac{\hbar}{J_{0}(\eta_{M})\alpha}\sqrt{\omega_{M}^{2}-\Omega_{0}^{2}}}{\frac{mg}{\hbar}}.
\]
 Hence, the condition required for having the avoided crossings sufficiently
separated, and, consequently, for soundly applying the LZ model at
each crossing is given by }

\textcolor{black}{
\[
\tau_{LZ}\ll\tau_{CC}.
\]
Using the expressions of $\tau_{LZ}$ and $\tau_{CC}$ and taking
$J_{0}(\eta_{M})\sim1$ and $J_{1}(\frac{\Omega_{M}}{\omega_{M}})\sim\frac{\Omega_{M}}{\omega_{M}}$
for the magnitude of the Bessel functions, the above restriction is
rewritten as 
\[
\Omega_{M}\ll\sqrt{\omega_{M}^{2}-\Omega_{0}^{2}}
\]
}

\textcolor{black}{vi) The probability of transition between adiabatic
states, $P_{LZ}^{(a)}$, is obtained by replaci}ng, in the expression
given by the standard LZ model \citep{key-Landau,key-Zener}, the
parameters $v$ and $\zeta$ by their counterparts in the present
system, $v_{eff}$ and $\zeta_{eff}$. Accordingly, we can write 
\begin{equation}
P_{LZ}^{(a)}=e^{-2\pi\left|\zeta_{eff}\right|^{2}/v_{eff}}.
\end{equation}
 The validity of the whole approach, and, in particular, of this expression
is confirmed by its applicability to the reproduction of the experimental
results on interferometry.

vii) The effects of including a nonzero detuning $\delta\neq0$ in
the SOC realization can be readily evaluated. The detuning is incorporated
into our approach by replacing $\alpha q$ by $\frac{\hbar\delta}{2}+\alpha q$.
Consequently, the \emph{parameter} $c(q)$ is shifted as $c(q)=J_{0}(\eta_{M})(\alpha q+\frac{\hbar\delta}{2})$.
\textcolor{black}{Also shifted are the functions $d(q)$ and $e(q)$.
In order to analyze how those shifts affect the effective LZ parameters
$v_{eff}$ and $\zeta_{eff}$, the following remarks are pertinent.
First, one must take into account that $d(q)$ and $e(q)$ incorporate
the dependence on $\delta$ through $c(q)$; in particular, $d(q)$
is proportional to $c(q)$. Second, the crossing, which is reached
when $e(q)$ takes a zero value, occurs at a value of the function
$c(q)$, let us call it $c(q_{C})$, which does not depend on $\delta$:
as can be seen from Eq. (20), $c(q_{C})$ is merely determined by
$\omega_{M}$ and $b$. As a consequence, no dependence on $\delta$
is apparent in $d(q_{C})$, and, in turn, one can conclude that the
coupling parameter of the LZ model $\zeta_{eff}=\frac{1}{\hbar}\frac{d(q_{C})}{2}$
is not affected by $\delta$. No changes are either observed in $v_{eff}$.
Actually, a nonzero detuning simply alters the quasimomentum at the
crossing $q_{C}$, and, consequently, leads to the appearance of a
constant term in the linear variation of the energy mismatch, which
does not modify the probability of transition.}

\section{Concluding remarks}

The study provides a logical framework where the experimental results
can be understood and the possibility of implementing controlled variations
of the observed dynamics can be contemplated. The operativity of the
study is rooted in the compact character of the model. Indeed, it
is the robustness of the LZ scenario that allows the simplified presentation
of the dynamics. Two examples are pertinent. First, since the activation
of the transitions takes place only near the crossings, it is possible
to cast the probability of transition into the standard LZ form by
taking the coupling parameter at the quasimomentum value of the crossing
point. Second, the applicability to the scenario with two avoided
crossings is also associated to the soundness of the LZ model: the
validity is guaranteed provided that the crossings are sufficiently
separated for assuming that approximate\emph{ asymptotic-regime} conditions
are reached between them.

By revealing the mechanisms responsible for the observed effects,
the study provides us with predictive power on the feasibility of
controlling diverse extensions of the setup. Particularly useful to
the design of strategies of control is to have the precise functional
form of the coupling term present in the emergent LZ model. Indeed,
we have seen that it is the specific form of the coupling term that
leads to the differential role of a nonzero detuning in the modulated
system. Another interesting question that can be raised in connection
with the coupling characteristics is the possibility of inhibiting
the interband transitions by appropriately choosing the modulation
parameters. Specifically, given the form of $d(q)$, one can speculate
on the feasibility of canceling the coupling by adjusting the amplitude
$\Omega_{M}$ and the frequency $\omega_{M}$ of the applied sinusoidal
driving in order to procure a zero of the factor $J_{1}(\eta_{M})$.
In fact, the analysis of that proposal cannot be made in the applied
theoretical framework. In this sense, we must recall a crucial step
in our procedure: the Hamiltonian with the multiple driving terms
given by Eq. (\ref{eq:DevelopDrivenH}) was simplified using arguments
relative to the magnitude of the ordinary Bessel functions. Namely,
$\eta_{M}$ was assumed to be small enough for allowing us to safely
neglect the contributions of terms associated to Bessel functions
of order higher than $n=1$. Consequently, the realization of an argument
$\eta_{M}$ close to a zero of $J_{1}(\eta_{M})$, which is outside
the previously assumed working range, demands a reevaluation of the
applied reduction. Actually, it is trivially shown that around the
first zero of $J_{1}(\eta_{M})$, different higher-order oscillating
terms present contributions of similar magnitude. Hence, a simple
reduction of the setup is not feasible.

Also important are the predictions allowed by our work on the potential
use of different acceleration mechanisms. Since the transitions have
been shown to be basically determined by the crossing-point properties,
a system response with characteristics similar to those found in the
study can be expected for alternative acceleration schemes. The case
of a variation of the quasimomentum induced by a harmonic trapping
must indeed be tackled in standard implementations. Although one can
assume that the potential interband transitions can be described using
a formalism similar to that used for gravitational acceleration, the
analytical description of such scenario seems challenging given the
nontrivial dynamics between transitions and the periodicity of the
evolution. Correspondingly, the analysis of interferometry results
when the harmonic-trap effects are important can be expected to be
complex. \textcolor{black}{Also interesting is to consider the potential
applicability of the presented method to confinement in an accelerated
optical lattice. Actually, the acceleration of an optical lattice
implies including in the Hamiltonian a term completely analogous to
the gravitational potential considered in the present study. However,
from the periodic character of the lattice potential, which implies
working with the Bloch-state representation, a variety of differential
effects can be expected. Indeed, as $\vec{P}$ is not a constant of
the motion in the confined system, we conjecture that a much more
complex treatment is needed to analyze the system dynamics in each
period of the lattice. }

Finally, a comment on the single-particle approach applied in the
analysis is pertinent. In \citep{key-Llorente2}, where the same atomic
system was studied, a detailed analysis of nonlinear effects was presented.
It was shown that, given the characteristics of the atomic-interaction
strengths, it can be assumed that the many-body dynamics do not significantly
modify the outcome of the single-approach picture. Specifically, the
general predictions of the model were found to be robust when atomic-interaction
effects were incorporated into the description. Those conclusions
are in the line of the operative scheme used in the interpretation
of the experimental work, where a single-particle approach was found
to be sufficiently accurate to give a satisfactory understanding of
the measurements. In this respect, it is also relevant to take into
account the recent study presented in \citep{key-TunableNonlLZ},
where it is reported that a significant modification of the experimental
conditions is required for making the interaction effects to be relevant
to the performance of the LZ model.

\end{document}